\documentclass[twocolumn,showpacs,preprintnumbers,amsmath,amssymb,pre]{revtex4}
\usepackage{graphicx}

\begin{document}

\title{Gibbs attractor: a chaotic nearly Hamiltonian system,
driven by external harmonic force}
\author{P. V. Elyutin}

\email{pve@shg.phys.msu.su}
\affiliation {Department of Physics,
Moscow State University, Moscow 119992, Russia}

\date{8 September, 2003}

\begin{abstract}
A chaotic autonomous Hamiltonian systems, perturbed by small
damping and small external force, harmonically dependent on time,
can acquire a strange attractor with properties similar to that of
the canonical distribution - the Gibbs attractor.  The evolution
of the energy in such systems can be described as the energy
diffusion.  For the nonlinear Pullen - Edmonds oscillator with two
degrees of freedom the properties of the Gibbs attractor and their
dependence on parameters of the perturbation are studied both
analytically and numerically.
\end{abstract}
 \maketitle

\vspace{25mm}

\section{Introduction}

\vspace*{-1mm}The Brownian motion of a particle of mass $m$ in the
static potential $U(\mathbf{r})$ can be described by the system of
Langevin equations
\begin{equation}\label{1}
m\ddot {\mathbf{r}}+\gamma \dot {\mathbf{r}}+\nabla U\left(
{\mathbf{r}} \right)={\mathbf{f}} \left( t \right),
\end{equation}
where $\gamma$ is the viscous damping parameter and
${\mathbf{f}}(t)$ is the random force (white noise) with Gaussian
distribution and mean values
\begin{equation}\label{2}
\left\langle {f_i\left( t \right)} \right\rangle
=0,\,\,\,\,\,\,\,\left\langle {f_i\left( t \right),f_j\left( {t'}
\right)} \right\rangle =2D\delta _{ij}\delta \left( {t-t'}
\right);
\end{equation}
here $\delta_{ij}$ is the Kronecker symbol and $\delta(x)$ is the
Dirac delta-function \cite{K75}. The damping term and the random
force provide a phenomenological description of the interaction
of the particle (in the potential) with its environment, that is
frequently called a heat bath.  The dynamical system determined
by Eqs. (\ref{1}) with $\gamma =0$ and $D=0$ will be called
isolated.

The ensemble of Brownian particles eventually will come to an
equilibrium with the stationary probability distribution in the
phase space given by the equation
\begin{equation}\label{3}
Q\left( {\mathbf{p},\mathbf{r}} \right)=N\exp \,\left(
{-{{H\left( {\mathbf{p},\mathbf{r}} \right)} \over \Theta }}
\right),
\end{equation}
where ${\mathbf{p}}=m\dot{\mathbf{r}}$ is the particle momentum,
$H\left( {\mathbf{p},\mathbf{r}} \right)={{\mathbf{p}^2} /
{2m}}+U\left( \mathbf{r} \right)$ is the Hamiltonian function of
the isolated system, $ \Theta ={D / \gamma }$ is the temperature,
and $N$ is the normalization constant.  The distribution Eq.
(\ref{3}) is known as the canonical, or Gibbs, distribution; it
serves as a central point of the equilibrium statistical physics
\cite{LL95}.  From Eq. (\ref{3}) follows the equilibrium energy
distribution,
\begin{equation}\label{4}
Q\left( E \right)=N\Phi \left( E \right)\exp \left( {-{E \over
\Theta }} \right),
\end{equation}
where
\begin{equation}\label{5}
\Phi \left( E \right)=\int {\delta \left( {E-H\left(
{\mathbf{p},\mathbf{r}} \right)} \right)d\mathbf{p}d\mathbf{r}}
\end{equation}
is the energy density of the phase volume on the energy surface
$H\left( {\mathbf{p},\mathbf{r}} \right)=E$.

The canonical distribution Eq. (\ref{3}) holds for the equilibrium
state of the system Eq. (\ref{1}) irrespectively of
dimensionality of the configuration space $d$ and regardless of
the nature of motion of the isolated system, be it regular
(periodic, quasiperiodic) or chaotic.

Let's replace the random force ${\mathbf f}(t)$ by a regular one
that depends on time harmonically; the equations of motion will
take the form
\begin{equation}\label{6}
m\ddot {\mathbf{r}}+\gamma \dot {\mathbf{r}}+\nabla U\left(
{\mathbf{r}} \right)={\mathbf{F}} \sin \omega t .
\end{equation}
If the motion of the isolated system is strongly chaotic, that
is, nearly ergodic on the energy surfaces in a wide range of
energy values, then we can expect that the motion of weakly
perturbed system ($\gamma$ and $F$ are small) will be chaotic
too.  In parallel with the physical picture of the Brownian motion
given above, the external force ${\mathbf{F}} \sin \omega t $
will slowly change the energy of the system, whereas the damping
$\gamma$ will provide a sink for the excessive energy, thus
creating a possibility for the equilibrium.

The dissipative non-autonomous system Eq. (\ref{6}) may
demonstrate the chaotic motion on a strange attractor, that has
much in common with the canonical distribution Eq. (\ref{3}). For
this reason it will be called the Gibbs attractor.

The main purpose of this paper is to give an example of a system
with the Gibbs attractor and to describe its main features.  They
mostly depend on the kinetics of energy exchange between the
perturbation and the isolated system, that can be described as a
process of energy diffusion.

The problem that we have formulated is on a crossroad of several
lines of research in nonlinear dynamics and non-stationary
statistical physics.  First, it is linked to the theory of chaos
in non-autonomous Hamiltonian systems.  In this theory the
concept of energy (or action) diffusion is used for description
of the infinite chaotic motion above the threshold of its onset -
in the models like periodically kicked rotator \cite{Ch79} or a
hydrogen atom in the microwave field \cite{DKSh83,CCh+87}. Unlike
our problem, in these models the strong periodic perturbation is
the source that thrusts chaos on the system. Secondly, our
problem is related to the theory of energy absorption by chaotic
systems with parametric modulation, that is usually developed
(both in quantum and classical approaches) for models of
billiards with varying form \cite{J93,C99}.  As opposed to our
problem, the main concern here is the study of slow variations,
whereas we are interested in the perturbation with frequencies
that are comparable to the characteristic frequencies of the
unperturbed systems.  Furthermore, in both aforementioned
theories the damping is completely neglected.  Thirdly, our
problem has relation to the actively developing theory of the
Brownian motion under the influence of colored noise
\cite{MAR00,QH00,MA+01,B01}, since one may consider the harmonic
perturbation as a limiting (monochromatic) case of the
narrow-band noise.  In contrast to our problem, in this theory
the unperturbed motion is mainly one-dimensional, thus regular,
and damping is considered strong.  Lastly, we note a connection
to the issue of the effects of weak noise and damping on the
Hamiltonian systems, that was discussed recently in the context
of the problem of decay of metastable chaotic states
\cite{PK99,PFE02}.

The rest of the text is organized as follows.  In Sec. II the
main equation for the energy diffusion is deduced in two ways.
Sec. III consists of the description of the unperturbed model,
the Pullen - Edmonds nonlinear oscillator, and main features of
its chaotic motion.  Sec. IV comprises the description of main
properties of the Gibbs attractor in the model - its limits in
the phase space, energy distribution, conditions of existence and
character of the energy correlation.  Sec. V contains the
concluding remarks.

\section{Energy diffusion}

We will study the evolution of the distribution of energy values
$Q(E,t)$ for the Hamiltonian chaotic system, perturbed by small
damping and small harmonic force.  For the derivation of the
equation of evolution for $Q(E,t)$ we at first neglect the damping
(we shall restore it later, in Subsec. C).

\subsection{The quantum approach}

The simplest approach is to start from the quantum model of the
unperturbed system. Let's assume that at the initial moment the
system is in a stationary state $|n\rangle$ with the energy $E$.
The external harmonic force will induce transitions with
absorption (+) and emission ($-$) of the quanta $\hbar \omega$.
If the motion of the classical system is chaotic and the power
spectrum of the active coordinate $x$ is continuous, then for its
quantum counterpart in the quasiclassical case the density of
final states with allowed transitions is high, and the rate of
these transitions can be described by the Fermi golden rule. With
the account of dependence of matrix elements and density of states
on the energy within the transition range, we can obtain for the
rate of transitions the expression
\begin{equation}\label{7}
\dot W_\pm \left( E \right)={{2\pi } \over {\hbar ^2}}\cdot
{{F^2} \over 4}\left[ {S\pm {{\hbar \omega } \over 2}\left(
{S'+S{{\rho '} \over \rho }} \right)} \right],
\end{equation}
where $S=S_x\left( {E,\omega } \right)$ is the power spectrum of
the coordinate $x$ and $\rho =\rho \left( E \right)$ is the
density of states of the isolated system, both taken at the
energy $E$, and dashes mean differentiating in energy
\cite{ESh96}.

The resonant absorption and emission of quanta populate narrow
bands of levels, that are located on the energy scale around
values $E'=E \pm k \hbar \omega$ with integer $k$. We denote the
probability of finding the system in such band of states as
$Q(E')$. With the account of one-photon transitions we can write
the balance equation
\begin{eqnarray}\label{8}
{{dQ\left( E \right)} \over {dt}}=-Q\left( E \right)\left( {\dot
W_++\dot W_-} \right)\\
\nonumber\qquad +Q\left( {E+\hbar \omega } \right)\dot W_++Q\left(
{E-\hbar \omega } \right)\dot W_-.
\end{eqnarray}

Assuming $Q(E,t)$ to be a smooth function of $E$, we can expand
the arguments in the second and third terms in the RHS of Eq.
(\ref{8}) to the order of $\hbar^2$ inclusive.  This yields the
equation in partial derivatives, that describes the energy
diffusion in the Hamiltonian system under the influence of the
external harmonic field:
\begin{equation}\label{9}
{{\partial Q} \over {\partial t}}+{\partial  \over {\partial
E}}\left( {AQ} \right)-{\partial \over {\partial E}}\left(D
{{\partial Q}\over {\partial E}} \right)=0.
\end{equation}
Here the coefficients of drift $A$ and of diffusion $D$ are given
by the expressions
\begin{equation}\label{10}
A(E, \omega)={\pi  \over 2}\omega ^2F^2S{{\rho '} \over \rho
},\,\,\,\,\,\,\,\,\,\,\,\,\,D(E, \omega)={\pi  \over 2}\omega
^2F^2S.
\end{equation}
It is essential that in the quasiclassical case, when the density
of states could be expressed by the Weyl formula
\begin{equation}\label{11}
\rho \left( E \right)={{\Phi \left( E \right)} \over {\left(
{2\pi \hbar } \right)^d}},
\end{equation}
the coefficients $A$ and $D$ do not depend on Planck's constant
$\hbar$.  The analogous derivation of the equation for the energy
diffusion in the conservative one-dimensional system under the
influence of white noise was used in \cite{ER01}.

\subsection{The classical approach}
The equation Eq. (\ref{9}) does not depend on Planck's constant,
but the condition of its applicability does.  The Fermi golden
rule is based on the assumption of the resonant character of the
transitions, which is justified only when the transition rate
$\dot W$ is much smaller then the perturbation frequency $\omega$.
Thus the diffusion coefficient must obey the strong inequality
\begin{equation}\label{12}
D \ll \hbar^2 \omega^3,
\end{equation}
that is too restrictive for small $\hbar$.

However, we can rederive Eq. (\ref{9}) classically.  In the zeroth
approximation we neglect the influence of the external force on
the law of motion $x(t)$.  Then the instantaneous rate of the
energy change is $\dot E = \dot x(t)F(t)$.  Variation of the
energy (for the time interval $T$), $\Delta(T)$, in this
approximation vanishes on the average for symmetry reasons,
$\langle\Delta(T)\rangle=0$, whereas its averaged square is
\begin{equation}\label{13}
\langle\Delta ^2\left( T \right)\rangle=\int\limits_0^T
{\int\limits_0^T {B_v\left( {t_1-t_2} \right)F\left( {t_1}
\right)F\left( {t_2} \right)dt_1dt_2}},
\end{equation}
where $B_v(\tau)$ is the correlation function of the
$x$-component of velocity.  If the unperturbed motion is ergodic,
then the correlation function is determined by the microcanonical
average
\begin{equation}\label{14}
B_v\left( \tau  \right)={1 \over {\Phi \left( E \right)}}\int
{\dot x\left( 0 \right)\dot x\left( \tau  \right)}\delta \left(
{E-H\left( {\mathbf{p},\mathbf{r}} \right)}
\right)d{\mathbf{p}}\,d{\mathbf{r}},
\end{equation}
where $x(0)$ and $x(\tau)$ are taken on the trajectory that
starts at $t=0$ at the phase point $\{\mathbf{p},\mathbf{r}\}$.
For times $T$ much larger than the time of decay of velocity
correlations $\tau_c$, we can rewrite Eq. (\ref{14}) in the form
\begin{equation}\label{15}
\langle\Delta ^2\left( T \right)\rangle={{F^2} \over
2}\int\limits_0^T {d\theta \int\limits_{-\infty }^\infty
{B_v\left( \tau \right)}}\cos \omega \tau d\tau.
\end{equation}
The internal integral is proportional to the power spectrum of
velocity, $S_v\left( {E,\omega } \right)$.  Since  $S_v\left(
{E,\omega } \right)=\omega ^2S_x\left( {E,\omega } \right)$, we
obtain for the coefficient of energy diffusion the expression
$D\left( {E,\omega } \right)=({\pi  /2})\omega ^2F^2S_x\left(
{E,\omega } \right)$, that coincides with the one given above in
Eq. (\ref{10}).

Let's assume that in the initial state the system has the energy
$E_0$ and denote $Q_0(E,t)$ the energy distribution with this
initial condition, $Q_0(E,0)=\delta(E-E_0)$.  The equation Eq.
(\ref{15}) can be rewritten as
\begin{equation}\label{16}
\left\langle {\Delta ^2\left( T \right)} \right\rangle
=2\int\limits_0^T {D\left( {E_0} \right)d\theta }.
\end{equation}
The distribution $Q_0$ spreads with time, including energy
surfaces with different densities $\Phi$ and different diffusion
coefficients $D$.  In the first approximation we can take this
into account by averaging of these functions with the evolving
distribution $Q_0(E,t)$:
\begin{equation}\label{17}
\left\langle {\Delta ^2\left( t \right)} \right\rangle
=2\int\limits_0^T {{{d\theta } \over {\Phi \left( {E_0}
\right)}}\int {dEQ_0\left( {E,\theta } \right)}}\,G\left( E
\right)
\end{equation}
where $G\left( E \right)=\Phi \left( E \right)D\left( E \right)$.
In the time interval in which the distribution $Q_0(E,t)$ can be
considered narrow, the function $G(E)$ can be expanded in the
Taylor series to the first order in $\Delta = E-E_0$.  Thus we
come to the equation
\begin{equation}\label{18}
\left\langle {\Delta ^2\left( t \right)} \right\rangle
=2\int\limits_0^T {\left( {D(E_0)+{{G'(E_0)} \over {\Phi \left(
{E_0} \right)}}\left\langle {\Delta \left( \theta  \right)}
\right\rangle } \right)}d\theta.
\end{equation}
If the system on the average changes its energy with a constant
rate, $\left\langle {\Delta \left( t \right)} \right\rangle
=\alpha t$, then $\left\langle {\Delta ^2\left( t \right)}
\right\rangle =\alpha ^2t^2+2Dt$.  Substituting this expressions
in Eq. (\ref{18}), we obtain
\begin{equation}\label{19}
D=D(E_0),\,\,\,\,\,\,\,\,\,\alpha ={{G'\left( {E_0} \right)} \over
{\Phi \left( {E_0} \right)}}.
\end{equation}
The average rate of the energy variation in the state with given
distribution $Q(E,t)$  can be obtained from the equation Eq.
(\ref{9}):
\begin{equation}\label{20}
\langle \dot E\rangle = \int {\left( {A+D'} \right)}Q\left( {E,t}
\right)dE.
\end{equation}
From Eqs. (\ref{19}) and (\ref{20}) for the drift coefficient we
obtain the expression $A\left( {E,\omega } \right)=D\left(
{E,\omega } \right)({{\Phi '\left( E \right)} / {\Phi \left( E
\right)}})$, that coincides with the one given above in Eq.
(\ref{10}), since the density of states is proportional to the
density of the phase volume (see Eq. (\ref{11})).

In the classical derivation of Eq. (\ref{20}) we have assumed that
the spreading of the initially localized energy distribution
during the correlation time $\tau_c$ is such that the corrections
to the diffusion coefficient are small in comparison with its zero
order value, $ \sqrt {D\tau _c}D'<<D$. Estimating the derivative
as $D' \sim D/E$, where $E$ is a characteristic energy of the
system, we obtain the condition of applicability of the classical
equation of energy diffusion,
\begin{equation}\label{21}
D\ll{{E^2} \over {\tau _c}},
\end{equation}
which is much more tolerant than Eq. (\ref{12}).  Bridging the gap
between Eq. (\ref{12}) and Eq. (\ref{21}) remains a challenge for
the quantum theory.

\subsection{The account of damping}
Now we turn to the account of damping.  For systems in which the
logarithmic rate of energy damping is constant, $\dot E = -\gamma
E$, the energy distribution could be written as
\begin{equation}\label{22}
Q\left( {E,t} \right)=e^{\gamma t}\psi \left( {Ee^{\gamma t}}
\right),
\end{equation}
where $\psi(z)$ is an arbitrary positive integrable function. This
functional form can be expressed by the equation in partial
derivatives
\begin{equation}\label{23}
{{\partial Q} \over {\partial t}}=\gamma {\partial  \over
{\partial E}}\left( {EQ} \right).
\end{equation}
By combining Eqs. (\ref{9}) and (\ref{23}) we obtain the equation
of energy diffusion in the perturbed chaotic system with damping:
\begin{equation}\label{24}
{{\partial Q} \over {\partial t}}-{\partial  \over {\partial
E}}\left( {\gamma EQ-AQ+D{{\partial Q} \over {\partial E}}}
\right)=0.
\end{equation}
Its stationary solution is given by the formulas
\begin{eqnarray}
\label{25}Q(E)&=&N \Phi \left( E \right)\exp \left[ {R\left( E,
\omega \right)} \right],\\
\label{26}R\left( E, \omega \right)&=&-\gamma \int\limits_{0}^E
{{\varepsilon  \over {D(\varepsilon ,\omega )}}}d\varepsilon,
\end{eqnarray}
where $N$ is the normalization constant.   The stationary
distribution could be canonical, Eq. (\ref{6}), only in the
special case when the diffusion coefficient is exactly
proportional to energy.

\section{The model}
The equations Eqs. (\ref{25},\ref{26}) could be checked by direct
numerical solution of the system Eq. (\ref{6}) for any given
potential $U(\mathbf{r})$. The calculation of $Q(E)$ is relatively
easy; however, the estimation of the integral in the RHS of Eq.
(\ref{26}) is much more demanding, since at every integration step
in $\varepsilon$ the power spectrum $S_x(\varepsilon, \omega)$ has
to be calculated with sufficient accuracy.  Instead of this
computation we will proceed with analytical approaches workable
with a special choice of the model.

We take the Pullen - Edmonds oscillator \cite{PE81}, that
describes the two-dimensional motion of a particle in the quartic
potential, as the model for the isolated system.  The Hamiltonian
of this system is
\begin{equation}\label{27}
H={1 \over {2m}}\left( {p_x^2+p_y^2} \right)+{{m\omega _0^2}
\over 2}\left( {x^2+y^2+{{x^2y^2} \over {\lambda ^2}}} \right).
\end{equation}
In the following we use the particle mass $m$, the frequency of
small oscillations $\omega_0$, and the nonlinearity length
$\lambda$ as unit scales, and write all equations in
dimensionless form.

The properties of chaotic motion of the Pullen - Edmonds model
are thoroughly studied \cite{M86,VZ87,EK89}.  With the increase
of energy the system becomes more chaotic both in extensive (that
is characterized by the measure of the chaotic component
$\mu_s(E)$ on the surface of Poincare section) and in intensive
(that is measured by the magnitude of the Lyapunov exponent
$\sigma (E)$) aspects. For values of energy $E>2.1$ the measure
$\mu_s >0.5$, and chaos dominates in the phase space; for $E>5$
the chaotic motion of the system is approximately ergodic
\cite{M86}.

For the selected model Eqs. (\ref{6}) have the form
\begin{eqnarray}\label{28}
\nonumber \ddot x+\gamma \dot x+x\left( {1+y^2}
\right)&=&F\sin \omega t,\\
\ddot y+\gamma \dot y+y\left( {1+x^2} \right)&=&0;
\end{eqnarray}
the external force is chosen to be directed along the $OX$ axis.
It may be noted that it is sufficient to couple the external force
to only one dynamical variable, as opposed to the Langevin forces
given by Eq. (\ref{2}): in chaotic system the interaction between
vibrational modes will secure the redistribution of energy.

For a particle in a two-dimensional potential the value of
$\Phi(E)$ is proportional to the area $\phi(E)$ of the region that
is bound by the equipotential line $U(x,y)=E$, namely
$\Phi(E)=2\pi \phi(E)$. For the Pullen - Edmonds model
\begin{equation}\label{29}
\phi \left( E \right)=4 \int\limits_0^{\sqrt {2E}} {\sqrt
{{{2E-x^2} \over {1+x^2}}}dx}.
\end{equation}
This integral can be calculated analytically:
\begin{equation}\label{30}
\phi \left( E \right)=4\sqrt {2E+1}\left[ {\textsf{K}}\left(
{\sqrt {{{2E} \over {2E+1}}}} \right)-\textsf{E}\left( {\sqrt
{{{2E} \over {2E+1}}}} \right) \right],
\end{equation}
where $\textsf{K}(x)$ and $\textsf{E}(x)$ are complete elliptic
integrals of the first and second kinds correspondingly.  For $E
\gg 1$ this function has the asymptotic form
\begin{equation}\label{31}
\phi \left( E \right)\approx \sqrt {32E}\left[ {\ln \sqrt {32E}-1}
\right].
\end{equation}

The approximate expression for the power spectrum of coordinate
$x$ for the Pullen - Edmonds model was obtained in Ref.
\cite{EK89} from the assumption of ergodicity of motion and in
the "frozen frequencies" approximation:
\begin{equation}\label{32}
S_x\left( {E,\omega } \right)={2 \over {3\phi (E) }}\left(
{2E+1-\omega ^2} \right)^{{3 \mathord{\left/ {\vphantom {3 2}}
\right. \kern-\nulldelimiterspace} 2}}\omega ^{-2}\left( {\omega
^2-1} \right)^{{{-1} \mathord{\left/ {\vphantom {{-1} 2}} \right.
\kern-\nulldelimiterspace} 2}},
\end{equation}
for $1<\omega <\sqrt {2E+1}$ and $S_x\left( {E,\omega }
\right)=0$ outside of this range.  The comparison of this formula
with the numerically found spectrum is shown in Fig. 1.

\begin{figure}[!ht]
\includegraphics[width=0.9\columnwidth]{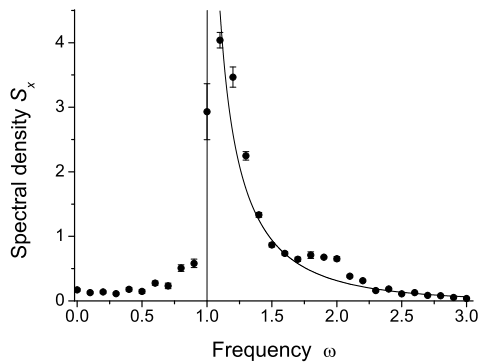}
\caption{\label{fig1} The dependence of the power spectrum of the
coordinate $S_x$ of the Pullen - Edmonds model Eq. (\ref{27}) on
frequency $\omega$ for the energy value $E=16$. Theoretical
estimate Eq. (\ref{32}) (line) and numerical calculation
(points).}
\end{figure}

It can be seen that the main body of spectrum, apart from a small
interval of frequencies around $\omega=1$, is described
satisfactorily.

\section{The Gibbs attractor}

In this section we study different properties of the motion on
the Gibbs attractor.  For the numerical calculation we will mainly
use the following set of parameters: $\gamma = 2\cdot10^{-3}$,
$F=0.3$ and $\omega=1.1$, that corresponds to $\Theta = 103$.
Later it will be referred to as a standard set.

\textbf{1.}  The motion of our model on the Gibbs attractor is
chaotic: numerical computation gives for its Lyapunov exponent
$\sigma = 0.65(2)$.  From the Kaplan - Yorke conjecture
\cite{KY79,LL92} it follows that the fractal dimensionality of
the attractor $D_F$ will differ from the dimensionality of the
phase space $d_p=4$ by a quantity of the order ${\gamma
\mathord{\left/ {\vphantom {\gamma  \sigma }} \right.
\kern-\nulldelimiterspace} \sigma }=3\cdot 10^{-3}$. This
difference is hardly noticeable: from the practical point of view
the Gibbs attractor densely fills the phase space, in resemblance
with the canonical distribution.

\textbf{2. } The energy distribution on the Gibbs attractor can be
determined by substitution of Eq. (\ref{32}) into Eq. (\ref{26});
thus we get
\begin{equation}\label{33}
R\left( E, \omega \right)=-{{3\gamma \sqrt {\omega ^2-1}} \over
{\pi F^2}}\int\limits_{}^E {{{\varepsilon\Phi \left( {\varepsilon}
\right)} \over {\left( {2\varepsilon+1-\omega ^2} \right)^{{3
\mathord{\left/ {\vphantom {3 2}} \right.
\kern-\nulldelimiterspace} 2}}}}d\varepsilon}.
\end{equation}

It is convenient to introduce the temperature parameter
\begin{equation}\label{34}
\Theta ={{\pi F^2} \over {3\gamma \sqrt {\omega ^2-1}}},
\end{equation}
that is similar to the temperature of the equilibrium created by
white noise heat bath and viscous damping, $\Theta = D/\gamma$: it
scales as the square of the force amplitude and as the inverse
value of the damping parameter.

For values of frequency $\omega \gtrsim 1$, after the
substitution of the asymptotic form Eq. (\ref{32}) in the
numerator and disregard of the quantity $(\omega^2-1)$ in the
denominator, the integration in Eq. (\ref{33}) yields the simple
formula:
\begin{equation}\label{35}
R\left( {E,\omega } \right)=-{E \over \Theta }\left( {\ln E+\ln
32-3} \right).
\end{equation}

In the derivation of Eq. (\ref{24}) we have assumed that the
energy damping in the absence of the external force is governed
by the equation $\dot E = -\gamma E$.  For the Pullen - Edmonds
model this relation must be corrected.  The exact equation for the
energy damping has the form $\dot E = - 2 \gamma T$, where $T$ is
the kinetic energy (properly averaged), but the virial ratio
$v(E)=2T/E$ in general case depends on the energy.  From the
virial theorem for the homogeneous potentials of the power $k$
\cite{LLI88} it follows that $v(E)={\mathrm{const}}=2k/(k+2)$. For
small energies, $E \ll 1$, the quartic term in the potential
$U(x,y)$ is negligible, and $v(E) \approx 1$, but in the domain $E
\gg 1$, that we are interested in, this term becomes important.
For purely quartic potential we have $v(E)=4/3$.  Thus we may
expect that for $E \gg 1$ in the Pullen - Edmonds model $1 < v(E)
< 4/3$; a naive interpolation leads to value $v(E)=7/6$. The
numerical calculation shows that the asymptotic value of $v(E)$
for high energies is very close to this value indeed, with the
accuracy not worse than 2\% for $E>20$. Therefore for very large
values of $\Theta$ we can improve the expression for $R(E,
\omega)$ just by multiplying it by the factor 7/6; the corrected
value of the damping parameter will be denoted as $\tilde \gamma$.

Comparison of the theoretical form of the probability distribution
$Q(E)$ (with the virial correction included) with its values found
numerically is shown in Fig. 2

\begin{figure}[!ht]
\includegraphics[width=0.9\columnwidth]{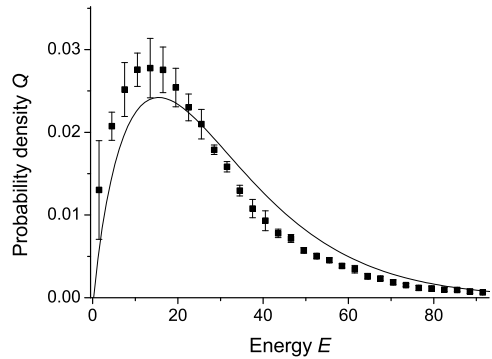}
\caption{\label{fig2} The stationary probability distribution
$Q(E)$ for the Gibbs attractor of the Pullen - Edmonds model with
the standard set of parameters ($\gamma = 2\cdot10^{-3}$, $F=0.3$
and $\omega=1.1$). Theoretical form given by Eqs. (\ref{25}),
(\ref{31}), and (\ref{35}) (line) and numerical calculation
(points).}
\end{figure}

The relaxation time $\tau _r$ for the diffusion process can be
estimated as the ratio of the square of the characteristic width
of the stationary distribution (e.g. the variance of the
distribution $V_E=\left\langle {E^2} \right\rangle -\left\langle E
\right\rangle ^2$) to the average value of the diffusion
coefficient: $\tau _r=V_E/\left\langle D \right\rangle $.  For the
standard set of parameters theoretical estimate based on Eqs.
(\ref{10}), (\ref{31}) and (\ref{32}) gives $V_E=420$,
$\left\langle D \right\rangle =0.67$ and $\tau _r=620$, whereas
the numerical calculation gives $V_E=315$, $\left\langle D
\right\rangle =0.42$ and $\tau _r=740$.  In our numerical
experiments we have used the time averaging over the intervals
about $1.7 \cdot 10^5 \approx 230 \tau _r$, that ensures high
accuracy of the stationary distribution.

\textbf{3.}  In opposition to the canonical distribution, the
Gibbs attractor is limited in the phase space. From the equation
for the energy dissipation the average rate of energy loss due to
damping in the state with the energy $E$ is
\begin{equation}\label{36}
P_-=\tilde\gamma E.
\end{equation}
The maximal rate of increase of the energy due to the influence of
the external force (for $F\ll 1$) is reached when the particle
moves along the $OX$ axis with its velocity in phase with the
external force, and equals to \vspace{-3mm}
\begin{equation}\label{37}
P_+=\sqrt {{E \over 2}}F.
\end{equation}
Balancing these quantities we find that the energy of the system
can not exceed the value
\begin{equation}\label{38}
E_+={{F^2} \over {2{\tilde \gamma} ^2}}.
\end{equation}
For the standard set $E_+=8.3\cdot 10^3=80\Theta $. The
probability to exceed this value with the energy distribution
given by Eqs. (\ref{25},\ref{26}) is about $\exp(-750)$; thus the
upper limit is not observable.

On the other hand, the attractor is not limited in the energy from
below.  The basic approximation of nearly ergodic motion is not
applicable in this range.  However, the system that at some moment
is at rest, in the state with $E=0$, can be driven by the external
force to the energies as high as
\begin{equation}\label{39}
E_0={{F^2\left( {3\omega ^2+1} \right)} \over {2\left( {\omega
^2-1} \right)^2}}
\end{equation}
(in harmonic approximation). For the standard set $E_0=4.72$. Thus
in our case the efficient communication between states with
negligibly small energies and those in the nearly ergodic domain
is possible.

\textbf{4.}   The motion of the system Eq. (\ref{28}) with the
standard set of parameters is found to be chaotic up to the
maximal available times of numerical calculation ($t=1.7 \cdot
10^5$ for 10 different initial conditions). The  time-averaged
value $\bar E$ of the energy of this motion is $\bar E \approx
26$.  Changes in parameters $\left\{ {\gamma ,F,\omega } \right\}$
that lead to diminution of the average energy of motion on the
attractor, such as the increase of damping, or lessening of the
amplitude of the harmonic force, or driving this force out of
resonance with the chaotic unperturbed motion, eventually suppress
the chaotic motion.  For some initial conditions the system, after
more or less prolonged interval of transient chaotic motion with
$\bar E \gg 1$ (we note in passing that main results of our theory
are applicable to long-living transient chaos as well as to the
perpetual one), rapidly turns to regular motion on a limit cycle
or a torus with $\bar E \lesssim 1$. This behaviour can have two
explanations.

(A)  The Gibbs attractor turns into a semiattracting set that
supports a metastable chaotic motion.

(B)  The Gibbs attractor endures, but its basin of attraction
shrinks, thus ceding larger parts of the phase space to basins of
regular attractors.

At present we can not choose between these alternatives.

We can establish the border of the domain of the apparent presence
of the Gibbs attractor conventionally, as the surface in the
parameter space at which for 80\% of randomly chosen initial
conditions the commencement of the regular motion takes less than
$3\cdot 10^4$ units of time. The detailed definition of this
surface is a laborious task, so we have restricted ourselves by
variation of each parameter in turn, with the other two being
fixed at the standard values.

With the increase of damping the domain of presence of the Gibbs
attractor is limited by the value $\gamma =5\cdot 10^{-3}$, where
the transient chaotic motion has $\bar E\approx 11$. With
diminishing $\gamma$ the chaos persists: even for $\gamma=0$ the
system displays chaotic motion, accompanied by unlimited spread of
the energy distribution with time.

With the decrease of the amplitude of the force the domain of
presence of the Gibbs attractor is limited by the value $F \approx
0.11$ where the transient chaotic motion has $\bar E\approx 6$.
The increase of the force $F$ up to values far beyond $F \approx
1$ does not bring any noticeable changes in the character of
motion.

This can be explained as follows. The magnitude of the external
force $F$ must be compared to the averaged (absolute) value of the
force in the same direction, created by the static potential,
$F_U=\left\langle {\left| {-{{\partial U} \mathord{\left/
{\vphantom {{\partial U} {\partial x}}} \right.
\kern-\nulldelimiterspace} {\partial x}}} \right|} \right\rangle
$.  On the energy surface $E$ it can be calculated by integration:
\begin{equation}\label{40}
F_U={4 \over {3\phi \left( E \right)}}\int\limits_0^{\sqrt {2E}}
{x{{3+2E-x^2} \over {1+x^2}}\sqrt {{{2E-x^2} \over {1+x^2}}}dx\,}.
\end{equation}
For large $E$ this expression has the asymptotics $F_U \sim E /
\ln E$.  The temperature parameter can be written as $\Theta =
\kappa F^2$ with $\kappa \gg 1$, and the average energy of chaotic
motion in the domain of existence of the Gibbs attractor, $\Theta
\gg 1$, depends on $\Theta$ as $\langle E \rangle \sim \Theta /
\ln \Theta$.  Thus for the average potential force we have the
estimate $\left\langle {F_U} \right\rangle \sim \kappa ^2F^2\ln
^{-2}\kappa F$, and for the ratio $ F / {\left\langle {F_U}
\right\rangle }$ we get
\begin{equation}\label{41}
{F \over {\left\langle {F_U} \right\rangle }}\sim {{\ln ^2\kappa
F} \over {\kappa ^2F}}.
\end{equation}
This quantity is always small and, moreover, decreases with growth
of $F$.  In physical terms, the harmonic force, albeit large in
comparison with the scales of the unperturbed model, heats the
system to the level of the energy content, at which it becomes
relatively small in comparison with the average potential force,
thus maintaining the energy diffusion picture adequate.

Finally, on the frequency scale the domain of presence of the
Gibbs attractor is limited by the band from $\omega \approx 0.93$
where the transient chaotic motion has $\bar E\approx 7$, to
$\omega \approx 1.6$ where $\bar E\approx 11$.

\textbf{5. } Since the energy of the system in our approach is
the main dynamical quantity, it is appropriate to study its
correlation function,
\begin{equation}\label{42}
B_E\left( \tau  \right)=\left\langle {E\left( 0 \right)E\left(
\tau  \right)} \right\rangle -\left\langle E \right\rangle ^2.
\end{equation}
The first term in the RHS of this equality could be expanded in
the Taylor series in the time shift $\tau$ and written as
\begin{equation}\label{43}
\left\langle {E\left( 0 \right)E\left( \tau  \right)}
\right\rangle =\sum\limits_{n=0}^\infty  {K_n{{\tau ^n} \over {n!
}}},
\end{equation}
where the coefficients $K_n$ are equal to the products of the
initial energy and of the initial value of energy's $n$-th time
derivative, averaged with the stationary distribution,
\begin{equation}\label{44}
K_n=\left\langle {E\left( 0 \right){{d^nE} \over {dt^n}}\left( 0
\right)} \right\rangle.
\end{equation}
From Eq. (24) for the local rate of the energy change we have
$\dot E=-\gamma E+A+D'$ and
\begin{equation}\label{45}
K_1=\left\langle {E\dot E} \right\rangle =\left\langle {-\gamma
E^2+AE+D'E} \right\rangle.
\end{equation}
The higher time derivatives of $E$ can be found by consequent
differentiation of Eq. (24) in time, recurrent substitutions and
integrations by parts.  For example, $K_2=\left\langle {E\ddot E}
\right\rangle $, where
\begin{equation}\label{46}
\ddot E=\left( {\gamma E-A-D'} \right)\left( {\gamma -A'-D''}
\right)+\left( {A''+D'''} \right)D.
\end{equation}
The normalized correlation function of energy $b_E\left( \tau
\right)={{B_E\left( \tau  \right)} \mathord{\left/ {\vphantom
{{B_E\left( \tau  \right)} {B_E\left( 0 \right)}}} \right.
\kern-\nulldelimiterspace} {B_E\left( 0 \right)}}$, found in the
numerical calculation, is compared to the theoretical estimate in
Fig. 3.

\begin{figure}[!ht]
\includegraphics[width=0.9\columnwidth]{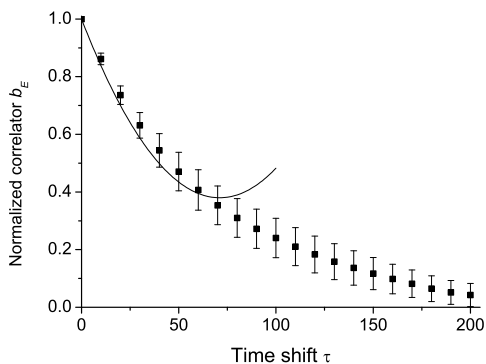}
\caption{\label{fig3} The normalized correlation function of
energy $b_E\left( \tau \right)$ for the motion on the Gibbs
attractor of the Pullen - Edmonds model with the standard set of
parameters ($\gamma = 2\cdot10^{-3}$, $F=0.3$ and $\omega=1.1$).
The time shift $\tau$ is measured in periods of harmonic field.
Theoretical approximation by first three terms of the expansion
Eq. (\ref{43}) (line) and numerical calculation (points).}
\end{figure}

The rate of decay of the energy correlations is determined by a
competition of the absorption, that dominates in the lower energy
range, and of the loss of energy through damping, that prevails in
the upper range; they are balanced by the stationarity equation
$\left\langle {\dot E} \right\rangle =0$. The dependence of this
rate on the parameters of the model is rather complicated.

\section{Conclusion}

The study of the system of the type Eq. (\ref{8}) may be
important, since for many experimentally relevant Hamiltonian
models one can indicate some mechanism of relaxation, that could
be approximated by the viscous damping terms at least
qualitatively.

The obvious candidates for the further studies of the Gibbs
attractors are chaotic billiards.  Our equations permit us to get
some simple estimates for this case.  Since a billiard
\textit{per se} has only two dimensional parameters, the particle
mass $m$ and some characteristic size $a$, the scale of time
depends on the initial conditions and is proportional to
$E^{-1/2}$.  Thus the power spectrum of the coordinate can be
written (with $m$ and $a$ as units) in the scaling form
\begin{equation}\label{47}
S_x\left( {E,\omega } \right)={1 \over {\sqrt E}}g\left(
{{\omega  \over {\sqrt E}}} \right),
\end{equation}
where $g(z)$ is some positive integrable function.  From
discontinuities of velocity for the law of motion $x(t)$  it
follows that the high frequency asymptotics of the power spectrum
has the form $S_x(E,\omega) \propto \omega^{-4}$. Combining this
formula with Eqs. (\ref{10}), (\ref{26}) and (\ref{47}),  we
obtain the approximate form of the exponent in the stationary
distribution:
\begin{equation}\label{48}
R\left( {E,\omega } \right)\approx -{{\sqrt E} \over \Theta },
\end{equation}
where $\Theta =cF^2\gamma ^{-1}\omega ^{-2}$ and $c$ is a
numerical constant that depends on the specific form of the
billiard.  This estimate is valid for $E \ll \omega^2$ for
billiards of any dimensionality.

\section*{ACKNOWLEDGMENTS}

This research was supported by the "Russian Scientific Schools"
program (grant \# NSh - 1909.2003.2).


\begin{thebibliography} {99}

\bibitem{K75}  V.I. Klyatzkin,  \textit{Statistical Description of the
Dynamical Systems with Fluctuating Parameters} (in Russian),
(Nauka, Moscow, 1975).

\bibitem{LL95}  L.D. Landau and E.M. Lifshitz,  \textit{Statistical
Physics. Part 1} (4th edition; in Russian), (Nauka, Moscow, 1995).

\bibitem{Ch79}  B.V. Chirikov, Phys. Reports  \textbf{52}, 263 (1979).

\bibitem{DKSh83} N.B. Delone, V.P. Krainov, and D.L. Shepelyansky,
Sov. Phys. Uspekhi \textbf{26}, 551 (1983). [Uspekhi Fiz. Nauk,
\textbf{140}, 355 (1983)].

\bibitem{CCh+87}  G. Casati, B.V. Chirikov, D.L. Shepelyansky, and I.
Guarnery, Phys. Reports  \textbf{154}, 77 (1987).

\bibitem{J93} C. Jarzynski, Phys. Rev. E  \textbf{48}, 4340 (1993).

\bibitem{C99}  D. Cohen, Phys. Rev. Lett. \textbf{82}, 4951
(1999); e-print cond-mat/9810395.

\bibitem{MAR00} R. Mankin, A. Ainsaar and E. Reiter, Phys. Rev. E
\textbf{61}, 6359 (2000).

\bibitem{QH00} J. Qiang and S. Habib, Phys. Rev. E \textbf{62},
 7430 (2000).

\bibitem{MA+01} R. Mankin, A. Ainsaar, A. Haljas, and E. Reiter,
Phys. Rev. E \textbf{63}, 041110 (2001)

\bibitem{B01} J.-D. Bao, Phys. Rev. E \textbf{63}, 061112 (2001).

\bibitem{PK99} I.V. Pogorelov and H.E. Kandrup, Phys. Rev. E
 \textbf{60}, 1567 (1999).

\bibitem{PFE02} T. Pohl, U. Feudel, and W. Ebeling, Phys. Rev. E
\textbf{65}, 046228 (2002).

\bibitem{PE81}  R.A. Pullen and A.R. Edmonds, J. Phys. A \textbf{14},
L477 (1981).

\bibitem{M86}  H.-D. Meyer,  J. Chem. Phys. \textbf{84}, 3147 (1986).

\bibitem{VZ87}  P.A. Vorobyev and G.M. Zaslavsky, Sov. Phys. JETP
\textbf{65}, 877 (1987) [ZhETF  \textbf{92}, 1564 (1987)].

\bibitem{EK89} P.V. Elyutin and V.G. Korolev, Mosc. Univ. Phys.
Bull. \textbf{44}, 106 (1989) [Vestn. Mosk. Univ. 3 Fiz. Astron.
\textbf{30}, 87 (1989)].

\bibitem{ESh96}  P.V. Elyutin and J. Shan, Phys. Rev. Lett. \textbf{77},
5043 (1996).

\bibitem{ER01} P.V. Elyutin and A.N. Rogovenko,  Phys. Rev. E  \textbf{63},
026610 (2001).

\bibitem{KY79} J.L. Kaplan and J.A. Yorke, in: \textit{Functional
differential equations and the approximations of fixed points},
eds. H.O. Peitgen and H.O. Walther, Lecture notes in mathematics,
Vol. 730, (Berlin, Springer, 1979), pp. 204-27.

\bibitem{LL92}  A.J. Lichtenberg and M.A. Lieberman, \textit{Regular
and Chaotic Dynamics} (Springer, Berlin, 1992).

\bibitem{LLI88} L.D. Landau and E.M. Lifshitz,  \textit{Mechanics}
(4-th edition; in Russian), (Nauka, Moscow, 1988), Sec. 10.



\end{thebibliography}
\end{document}